\documentclass[twocolumn,superscriptaddress,showpacs,preprintnumbers,amsmath,nofootinbib,amssymb]{revtex4}
\usepackage{epsfig}
\usepackage{graphicx}
\usepackage{dcolumn}
\usepackage{bm}
\usepackage{color}

\newcommand{\be}{\begin{equation}}
\newcommand{\ee}{\end{equation}}

\def\thebiblio#1{
\begin{center}\bf \large References
\end{center}
\list
{[\arabic{enumi}]}{\settowidth\labelwidth{#1.}\leftmargin\labelwidth
 \advance\leftmargin\labelsep
 \usecounter{enumi}}
 \def\newblock{\hskip .11em plus .33em minus -.07em}
 \sloppy
 \sfcode`\.=1000\relax}

\newcommand{\calr}{{\cal R}}
\newcommand{\calp}{{\cal P}}
\renewcommand{\[}{\begin{equation}}
\renewcommand{\]}{\end{equation}}

\begin{document}
%\begin{titlepage}

%\preprint{UGFT-223/08}

\preprint{KEK-Cosmo-44}
\preprint{KEK-TH-1387}

\title{The gravitino problem in supersymmetric warm inflation}

\author{Juan~C.~Bueno~S\'anchez}%
\email{jcbueno@fis.ucm.es}
\affiliation{%
Departamento de F\'isica At\'omica, Molecular y Nuclear, Universidad Complutense de Madrid, 28040 Madrid, Spain}
\affiliation{%
Department of Physics, University of Ioannina, Ioannina 45110, Greece}
\affiliation{%
Physics Department, Lancaster University, Lancaster LA1 4YB, UK}
\author{Mar~Bastero-Gil}%
\email{mbg@ugr.es}
\affiliation{%
Departamento de F\'{\i}sica Te\'orica y del Cosmos,
Universidad de Granada, Granada-18071, Spain}%
\author{Arjun~Berera}%
\email{ab@ph.ed.ac.uk}
\affiliation{%
SUPA, School of Physics and Astronomy,
University of Edinburgh, Edinburgh EH9 3JZ, UK}%
\author{Konstantinos~Dimopoulos}%
\email{k.dimopoulos1@lancaster.ac.uk}
\affiliation{%
Physics Department, Lancaster University, Lancaster LA1 4YB, UK}%

\author{Kazunori Kohri}%
\email{k.khori@lancaster.ac.uk}
\affiliation{%
Cosmophysics group, Theory Center, IPNS, KEK, Tsukuba 305-0801, Japan
}%
\affiliation{%
Department of Physics, Tohoku University, Sendai 980-8578, Japan
}%
\affiliation{%
Physics Department, Lancaster University, Lancaster LA1 4YB, UK}%

\date{\today}% It is always \today, today,
             % but any date may be explicitly specified
%\medskip

%\address{$^1$Physics Department, Lancaster University, Lancaster LA1 4YB, UK}
%\address{$^2$Departamento de F\'{\i}sica Te\'orica y del Cosmos,
%Universidad de Granada, Granada-18071, Spain}
%\address{$^3$School of Physics, University of Edinburgh,
%Edinburgh, EH9 3JZ, UK}

%\address{E-mail:j.buenosanchez@lancaster.ac.uk, mbg@ugr.es, ab@ph.ed.ac.uk,
%k.dimopoulos1@lancaster.ac.uk}

\begin{abstract}
The warm inflation paradigm considers the continuous production of
radiation during inflation due to dissipative effects. In its strong
dissipation limit, warm inflation gives way to a radiation dominated
Universe. High scale inflation then yields a high reheating
temperature, which then poses a severe gravitino overproduction
problem for the supersymmetric realisations of warm inflation. In this
paper we show that in certain class of supersymmetric models the
dissipative dynamics of the inflaton is such that the field can avoid
its complete decay after inflation. In some cases, the residual energy
density stored in the field oscillations may come to dominate over the
radiation bath at a later epoch. If the inflaton field finally decays
much later than the onset of the matter dominated phase, the
entropy produced in its decay may be sufficient to counteract the
excess of gravitinos produced during the last stages of warm inflation.
%We show that the survival of the inflaton field up to late times requires
%a vacuum expectation value for the field in the Planck scale. We also
%consider a generic supergravity model and show that the inflaton field
%decays completely after warm inflation. Although recently computed
%corrections to the thermal perturbation spectrum of warm inflation
%allow inflation at a very low scale, these corrections give rise to a
%blue tilted spectrum, currently disfavored in the absence of tensor
%perturbations.
\end{abstract}

%\end{titlepage}

\pacs{98.80.Cq, 11.30.Pb, 12.60.Jv}
\maketitle

%\tableofcontents

%\pagebreak

\section{Introduction}

The flatness required of inflationary potentials, to satisfy density
perturbation constraints, most commonly relies of Supersymmetry to
protect against radiative corrections.  Although SUSY is hugely
successful in building models of inflation, associated with it are
also some problems. One of these is the gravitino overproduction
problem. Very simply the problem is, ending inflation at too high a
temperature can lead to a gravitino abundance that is prohibited by
nucleosynthesis constraints when we consider a massive unstable
gravitino (in this paper we will mainly discuss the case of
the massive unstable gravitino with the mass of $m_{3/2} \sim
{\cal O}(1)$ TeV). One alternative is to end inflation at a lower
temperature, but there are also advantages to a high temperature exit
from inflation, notably it can lead to effective leptogenesis
\cite{Fukugita:1986hr,lepto}.

There are two distinct dynamical realizations of inflation,
cold and warm inflation.
Cold inflation is the standard scenario in which
the inflaton is assumed to be noninteracting during inflation,
and thus there is no particle production during inflation,
so the Universe supercools \cite{ci}.  Only after inflation, interactions
of the inflaton with other fields is assumed significant,
and a reheating phase occurs when the vacuum energy used
to drive inflation is converted into particles that forms
the subsequent radiation dominated phase.  In the alternative
warm inflation picture, the inflaton interacts with other fields
during the inflation phase, which leads to particle production
concurrent with inflationary expansion \cite{Berera:1995ie}
(for recent reviews please see
\cite{Berera:2008jn,Berera:2008ar,BasteroGil:2009ec}).
The presence of a radiation
energy density is not inconsistent with the General Relativity
requirements for realizing inflation, which only requires that the vacuum
energy dominates the energy density in the Universe.
Thus in the warm inflation picture, the presence of radiation during
inflation implies this phase smoothly ends into a radiation
dominated phase without a distinctively separate reheating
phase. This offers an alternative dynamic solution to the graceful
exit problem of inflation.  The presence of radiation then implies
the fluctuation of the inflaton, which are the primordial seeds
of density perturbations, are now thermal
\cite{im,Berera:1995wh,Berera:1999ws}, rather than the quantum
fluctuations that occur in cold inflation.

The equations of motion for the inflaton field and for the
radiation density in the presence of a dissipation mechanism are
\[\label{eq21}
\ddot{\phi}+3H(1+Q)\dot{\phi}+V'=0\,,
\]
and
\[\label{eq29}
\dot{\rho}_r+4H\rho_r=\Upsilon\dot{\phi}^2\,,
\]
where $Q\equiv \Upsilon/3H$, $\Upsilon$ is the dissipative
coefficient,  overdots stand for time derivative,
and $'\equiv\frac{d}{d\phi}$. Throughout the paper we use natural
units $c=\hbar=k_{\rm B}=1$ and Newton's gravitational constant is
\mbox{$8\pi G=m_P^{-2}$}, where $m_P=2.4\times10^{18}\,$GeV is the
reduced Planck mass.

Since interactions are important during warm inflation,
if the fields interacting with the inflaton are at high
temperature, then it is
difficult to control the thermal loop corrections
to the effective potential that is needed to maintain
the very flat potential required for
inflation \cite{Berera:1998gx,Yokoyama:1998ju}.
However, if the fields interacting with the inflaton
are at low temperature, then supersymmetry can be used
to cancel the quantum radiative corrections, and maintain a very
flat inflaton potential.
In this paper we consider a class of supersymmetric models for which
the dissipation coefficient $\Upsilon$ has been computed in the
equilibrium approach \cite{Berera:1998gx} for the low-temperature
regime \cite{Moss:2006gt,BasteroGil:2010pb}.
The dissipation mechanism is based on a
two-stage process \cite{br}.
The inflaton field
couples to heavy bosonic fields, $\chi$ and fermionic fields
$\psi_{\chi}$, which then decay to light degrees of freedom. These
light degrees of freedom thermalize to become radiation. The simplest
superpotential containing such an interaction structure is
\begin{equation}
W=%\frac{1}{3} \sqrt{\lambda} \Phi^3 +
g\Phi X^2+hXY^2, \label{W}
\end{equation}
where $\Phi$, $X$ and $Y$ denote superfields, and $\phi$, $\chi$
and $y$ refer to their bosonic components. Such an interaction
structure is common in many particle physics SUSY models during
inflation, the field $y$ and its fermionic partner $\bar{y}$
remain massless or very light, whereas the field $\chi$ and its fermion partner
$\psi_{\chi}$ obtain their masses through their couplings to
$\phi$, namely \mbox{$m_{\psi_{\chi}}=m_\chi=g\phi$}. The regime
of interest is when \mbox{$m_\chi,m_{\psi_{\chi}} > T > H$}, and
this defines what is usually referred to as the low-temperature
regime. For this regime the dissipation coefficient, when the
superfields $X$ and $Y$ are singlets, is found to be
\cite{Moss:2006gt,BasteroGil:2010pb}
\[\label{ups}
\Upsilon\simeq0.64\,g^2h^4\left(\frac{g\phi}{m_\chi}\right)^4
\frac{T^3}{m_\chi^2}\,,
\]
where $T$ is the temperature of the radiation bath,
\mbox{$\rho_r=C_rT^4$}, where $C_r=\pi^2 g_*/30$ and $g_*$ is the
number of relativistic degrees of freedom.
The above dissipative coefficient is calculated under the
adiabatic approximation, and associated with it are consistency
conditions which ensure that the microscopic dynamics is
faster than the macroscopic motion \cite{Berera:1998gx},
\[
\Gamma_{\chi  \ (\psi_{\chi})} \gg \frac{\dot \phi}{\phi}, H .
\]
In supersymmetric theories $C_r\simeq70$.  However, the
superfields $X$ and $Y$ may belong to large representations of a GUT
group. In that case, the dissipation coefficient picks up an extra
factor ${\cal N}={\cal N}_\chi {\cal N}^2_{\rm decay}$, where ${\cal
N}_\chi$ is the multiplicity of the $X$ superfield and ${\cal N}_{\rm
decay}$ is the number of decay channels available in $X$'s decay.
Typically, having enough dissipation during inflation requires large
multiplicites, ${\cal N}_\chi \sim  {\cal N}_{\rm decay}
\sim O(100)$. In that respect, string theory can be a natural place
for warm inflation due to the presence of large numbers of
moduli fields \cite{stringwi}.

Following this approach, it has been recently shown
\cite{BasteroGil:2006vr,Zhang:2009ge}
that chaotic and hybrid inflation models may
support some $50$ to $60$ $e$-foldings of warm inflation in the strong
dissipative regime, but such models can lead to
an overproduction of gravitinos \cite{Taylor:2000jw}.
In the context of hilltop models, it was found in
Ref.~\cite{BuenoSanchez:2008nc} that although warm inflation agrees
with current observations the temperature of the radiation at the end
of inflation exceeds the current bounds on thermal gravitino
production for massive gravitino with its mass $m_{3/2}={\cal O}({\rm
TeV})$.

In order to dilute the excess of gravitinos thermally produced towards
the end of warm inflation, it is possible to argue that the necessary
entropy production owes to the decay of the inflaton field, which
comes to dominate the Universe at a later epoch.  However, if the
ratio $Q\gg1$, the inflaton field decays completely right after
inflation and no later entropy production can be attributed to it. It
is worth emphasizing though, that $\Upsilon$, as given in
Eq.~(\ref{ups}), is time-dependent.  Moreover, it \textit{always}
falls faster than the Hubble parameter during the radiation dominated
epoch that follows after inflation: $\Upsilon\propto a^{-3}$ whereas
$H\propto a^{-2}$.  Therefore, if $Q$ is not too large, it is
plausible that soon after inflation the system moves into the weak
dissipation and hence the inflaton field does not decay completely
until a later epoch. Once in the weak dissipation regime, the average
scalar density of the field decreases as $a^{-3}$, hence it may
come to dominate over the radiation density thus driving a late
matter-dominated epoch. This epoch is terminated by the perturbative
decay of the inflaton field. The entropy produced by this decay may be
sufficient to dilute the excess of gravitinos.  After this, the
inflaton decays completely  before the Big-bang nucleosynthesis epoch
and the thermal bath of the hot big bang is recovered\footnote{When
we do not  consider the nonthermal gravitino production in
supergravity, the most conservative lower bounds on the reheating
temperature, $T_{R} > 3 -4$ MeV, comes from the insufficient
thermalisation of the neutrino background, which changes the $^{4}$He
abundance~\cite{RDcond}.}.

\section{Field evolution}\label{sec.evol}
In order to investigate the field evolution during inflation and
its subsequent oscillatory phase we consider a class of hilltop
models with the scalar potential given by
\[\label{eq26}
V=V_0f(\phi)\,,
\]
where $V_0$ is a characteristic density scale and $f$ is a
dimensionless function with a maximum at $\phi=0$ and a minimum at
$\phi=\phi_v$, where it vanishes. We parametrise the field's
expectation value by the dimensionless quantity $\delta$, defined
by
\[\label{eq24}
\phi\equiv\phi_v(1-\delta)\,.
\]
In the following we use subscripts ``$*$'', ``$e$'' and ``$o$'' to
denote the time when cosmological scales exit the horizon, the end
of inflation, and the onset of the field oscillations
respectively. We consider the two-stage dissipation mechanism
described in the introduction and take
\[
\Upsilon\simeq C_\phi T^3\phi^{-2}\,.
\]
where\footnote{This should be
small  in case of MSSM~\cite{Kamada:2009hy}.}  $C_\phi=0.64\,h^4{\cal N}$.

\subsection{Inflation and first reheating}
We are interested in the amount of expansion that follows after
the observable Universe exits the horizon, and so we require that
warm inflation is supported for around 50 $e$-foldings. In the
warm inflation paradigm, the slow-roll equations that apply during
inflation are
\[\label{eq37}
\dot{\phi}\simeq-\frac{V'}{3H(1+Q)}\quad\textrm{and}\quad\rho_r\simeq\frac{\Upsilon\dot{\phi}^2}{4H}\,.
\]
These equations hold for as long as the so-called \emph{modified}
slow-roll parameters, given by \cite{BasteroGil:2004tg}
\[
\epsilon=\frac{\epsilon_\phi}{(1+Q)}\,,
\]
\[
\eta=\frac{\eta_\phi}{(1+Q)} \,,
\]
and
\[
\epsilon_{HY}=\frac{1}{(1+Q)}\,\frac{V'}{3H^2}\frac{\Upsilon'}{\Upsilon}\,,
\]
are sufficiently small. In the above, $\epsilon_\phi$ and $\eta_\phi$
are the slow-roll parameters in the absence of a dissipative
mechanism
\[\label{eq43}
\epsilon_\phi=\frac{m_P^2}{2}\left(\frac{V'}{V}\right)^2\quad,\quad
\eta_\phi=m_P\frac{V''}{V}\,.
\]

Because the radiation density $\rho_r$ remains subdominant until
the end of inflation, we may obtain the evolution of $H$ during
inflation simply by integrating $d(H^2)\simeq\frac{dV}{3m_P^2}$.
Using Eq.~(\ref{eq26}), the Hubble parameter when cosmological
scales exit the horizon is
\[\label{eq23}
H_*\simeq H_0\,f_*^{1/2}\,,
\]
where $H_0$ is defined by $V_0\equiv3H_0^2m_P^2$. In the strong
dissipative regime of warm inflation, the accelerated expansion
lasts until the radiation density, constantly produced owing to
the dissipation mechanism, catches up with $V(\phi)$. Somewhat
before this time, the Hubble parameter slightly deviates from
$H\simeq H_0f^{1/2}$. If we neglect the kinetic density of the
field, which in the strong dissipation regime remains subdominant
during inflation, the Hubble parameter when the radiation density
catches up with $V(\phi)$ is well approximated by
\[\label{eq30}
H_e\simeq\sqrt{2}H_0\,f_e^{1/2}\,.
\]
Although the accelerated expansion finishes at this time, quasi-de
Sitter inflation ends somewhat earlier. Nevertheless, the amount
of expansion that follows from that moment until the end of
accelerated expansion is negligible. Therefore, we approximate the
number of $e$-foldings $N_*$ by considering that the exponential
expansion finishes at the time of potential-radiation equality.
Also at this time, the source term $\Upsilon\dot{\phi}^2$ in
Eq.~(\ref{eq29}) starts becoming subdominant, and the radiation
density separates from its attractor solution and starts scaling
as $\rho_r\propto a^{-4}$. On its part, and provided that $Q_e>1$,
the scalar field still obeys its slow-roll equation and continues
transforming its potential energy into radiation. Consequently,
immediately after the time of equality the scalar density becomes
subdominant and the Universe becomes radiation dominated. We
define the \emph{first reheating temperature} $T_{R_1}$ as the
temperature at the time of equality, namely
\[\label{eq39}
T_{R_1}\equiv T_e\simeq C_r^{-1/4}(V_0f_e)^{1/4}\,.
\]

\subsection{Onset of inflaton oscillations}
Because $\Upsilon$ depends on the temperature of the radiation, in
the radiation dominated phase $\Upsilon$ evolves in a manner
different than during inflation. As a result, the inflaton field
loses energy at a different rate. Here we compute the potential
density of the field when the field ceases to follow its attractor
equation. This time, owing to the dissipation term, is determined
by the condition
\[\label{eq19}
\Upsilon_o\sim|V''_o|^{1/2}\,.
\]
Integrating the slow-roll equation $\dot{\phi}\simeq|V'|/\Upsilon$
from the end of inflation until the onset of the fast-roll motion
we obtain the integral equation that determines $\phi_o$
\[\label{eq8}
I_\phi=\int_{\phi_e}^{\phi_o}\frac{d\phi}{V'\phi^2}\simeq-\int_{t_e}^{t_o}
\frac{dt}{C_\phi T^3}=I_t\,,
\]
where we used $\Upsilon=C_\phi T^3/\phi^2$. Using that in this
stage the Universe is radiation dominated, i.e. $H\simeq
H_e(a_e/a)^2$ and $T\simeq T_{R_1}(a_e/a)$, the integral $I_t$ may
be approximated by
\[\label{eq9}
I_t=-\int_{a_e}^{a_o}\frac{da}{(aH)C_\phi T^3}\simeq\frac1{5C_\phi
H_eT_{R_1}^3}\left[1-\left(\frac{a_o}{a_e}\right)^5\right]\,.
\]
The ratio $a_o/a_e$ can be estimated as follows. Using the definition of $\Upsilon$ and that $T\propto a^{-1}$
it is straightforward to obtain the relation
\[\label{eq31}
\Upsilon_o\simeq \Upsilon_e\left(
\frac{a_e}{a_o}\right)^3\frac{(1-\delta_e)^2}{(1-\delta_o)^2}\,,
\]
where $\Upsilon_e$ can obtained from the attractor equations and
$\Upsilon_o$ is determined by Eq.~(\ref{eq19}). The redshift
factor $(a_o/a_e)$ is then approximated by
\[\label{eq32}
\left(\frac{a_o}{a_e}\right)^5\sim\frac{
C_\phi^{20/21}(f\textrm{'}_e)^{10/7}(1-\delta_e)^{10/7}H_0^{10/21}
m_P^{25/21}}{2C_r^{5/7}|f\textrm{''}_o|^{5/6}
f_e^{5/14}(1-\delta_o)^{10/3}\phi_v^{5/3}}\,,
\]
where $\textrm{'}\equiv\frac{d}{d\delta}$. Substituting this expression into
Eq.~(\ref{eq9}) we can solve Eq.~(\ref{eq8}) numerically and find
$\delta_o$.

\subsection{Second reheating temperature}
In order to simplify our analytical treatment we assume in this
section that at $a=a_o$ the field starts performing fast
oscillations about its vev. In such case its average density
$\langle\rho_\phi\rangle$ can be readily computed
\cite{Turner:1983he}. Although this assumption naturally results in an
error estimating the scalar density of the field, in the Appendix
we explain how this can be corrected, thus obtaining an
accurate estimate of the average scalar density at late times.
Assuming then a sudden transition to the stage of fast
oscillations, the average density of the field is determined by
\[
\langle\dot{\rho}_\phi\rangle\simeq-3H(1+Q)\langle\rho_\phi\rangle\,.
\]

Although at the beginning of the oscillations it may be $Q>1$, we
keep the term $3H$ in the above as this becomes the dominant one
when the system moves into the weak dissipation regime. Using that
$H\simeq H_o\left(\frac{a_o}{a}\right)^2$ and that
$\langle\phi\rangle=\phi_v$, and hence $\Upsilon\simeq
\Upsilon_o\left(\frac{a_o}{a}\right)^3$, the average density of
the field can be readily obtained
\[\label{eq14}
\langle\rho_\phi\rangle\simeq(\rho_\phi)_o\left(\frac{a_o}{a}
\right)^3\exp\left\{-3\,Q_o\left(1-\frac{a_o}{a}\right)\right\}\,,
\]
where $Q_o$ can be expressed in terms of $\delta_e$ and $\delta_*$ after
using Eq.~(\ref{eq31}). From the above expression it follows
that $\langle\rho_\phi\rangle\propto a^{-3}$ a few Hubble times after
the onset of oscillations. Thus, provided that $Q_o$ is not large, the
inflaton may come to dominate the Universe at a later time well before
nucleosynthesis. We denote by  $t_{\rm eq}$ the time when the average
scalar density $\langle\rho_\phi\rangle$ catches up with $\rho_r$. The
subsequent matter-dominated phase must be sufficiently long so that
the entropy produced when the inflaton field decays completely at $t=t_{\rm dec}$
is enough to dilute the overproduction of
gravitinos. With the complete decay of the inflaton at $t=t_{\rm dec}$, which we
consider a free parameter, the
Hot Big Bang evolution is recovered. The second reheating temperature
$T_{R_2}$ is then
\[\label{eq15}
\,T_{R_2}\sim C_r^{-1/4}\langle\rho_\phi\rangle_{\rm
eq}^{1/4}\left(\frac{a_{\rm eq}}{a_{\rm dec}}\right)^{3/4}\,,
\]
which depends on $\delta_e$ and on the model parameters.
Obviously, to estimate this temperature we first need to have an
estimate for the quantities $\delta_*$ and $\delta_e$.

%%%%%%%%%%%%%%%%%%%%%%%%%%%%%%%%%%%%%%%%%%%%%%%%%%%%%%%%%%%%%%%%%%%%%%
\section{Observational constraints}\label{sec.obs}
%%%%%%%%%%%%%%%%%%%%%%%%%%%%%%%%%%%%%%%%%%%%%%%%%%%%%%%%%%%%%%%%%%%%%%

The purpose of this section is to determine $\delta_*$ and  $\delta_e$, employed to compute $T_{R_2}$, in terms of the model parameters. The \textit{raison d'\^etre} of this is to link the reheating temperatures $T_{R_2}$, computed according to Eq.~(\ref{eq15}), to observable quantities like the spectral index or its running, in turn determined by the model parameters. By doing this we manage to identify the interval of temperatures $T_{R_2}$ corresponding to certain range of values of the spectral index (see Fig.~\ref{fig2}).

The predicted amplitude of the curvature perturbation spectrum in
the strong dissipative regime of warm inflation is given by
\[
\calp_\calr\simeq\widetilde{\calp_\calr} \calp_\calr^{\rm (c)}
\]
where
\[
\widetilde{\calp_\calr}\simeq\left(\frac{3H^3}{2\pi
V'}\right)^2(1+Q)^{5/2}\frac{T}{H}\,
\]
has been known for sometime \cite{Berera:1999ws}, whereas the
correction $\calp_\calr^{(c)}$ was computed only recently
\cite{Graham:2009bf}. The correction term has a range of behavior,
depending on the details of the dynamics of the radiation energy density
component.  We have examined this range, but will not explore these
details here. The purpose of this paper is to highlight a general
mechanism for treating gravitino production within warm inflation,
and for that we will focus on one particular form of the correction
\cite{Graham:2009bf,mg}
for the dissipation coefficient
in Eq.~(\ref{ups}),
\[
{\calp}_\calr^{\rm (c)}\simeq
%\frac{1+3Q}{1+Q}(1+a_0Q^2+a_1Q^3)
\frac{1+a_0Q^2+a_1Q^3}{\sqrt{1+Q}}
\,,
\]
where $a_0\simeq0.662$ and $a_1\simeq3.26\times10^{-4}$.

Using Eqs.~(\ref{eq26}), (\ref{eq37}) and (\ref{eq23}), and
writing $Q$ in terms of $C_\phi$, $H_0$ and $\phi$ the above may be recast
as
\[\label{eq27}
\calp_\calr=\calp_\calr(C_\phi,H_0,\phi_*,\beta_i)\,,
\]
which can be solved numerically to determine $\phi_*$. Here $\beta_i$ denotes the rest of the model parameters which the amplitude of the spectrum may depend on.

In the strong dissipative regime warm inflation gives way to a
radiation dominated Universe. Neglecting the kinetic density of
the field, the end of the accelerated expansion occurs when
the radiation density catches up with the scalar potential, i.e.
when
\[
V_e=(\rho_r)_e\,.
\]
As mentioned earlier, the amount of quasi-de Sitter inflation is well
approximated by considering that this finishes at $\phi=\phi_e$.
To determine $\phi_e$ we rewrite the above condition expressing
$V_e$ in terms of $\delta_e$ and using the attractor equation for
$(\rho_r)$. The field value $\phi_e$ is then determined by
\[\label{eq28}
\frac{(f\textrm{'}_e)^{8/9}(1-\delta_e)^{8/9}}{f_e}\simeq\frac{
2^{11/9}C_{\phi }^{4/9}}{3^{1/9}
C_r^{1/3}} \left(\frac{H_0}{m_P}\right)^{2/9}\,.
\]

By simple inspection, it is clear that solving for $\phi_*$ and
$\phi_e$ from Eqs.~(\ref{eq27}) and (\ref{eq28}) is, in general, not amenable
to analytical treatment. Owing to this, we proceed alternatively by
solving $H_0$ and $C_\phi$ in terms of $\delta_*$ and $\delta_e$,
i.e.
\[\label{eq25}
H_0=H_0(\delta_*,\delta_e)\quad{\rm and}\quad
C_\phi=C_\phi(\delta_*,\delta_e)\,.
\]

The amount of exponential expansion that follows from horizon crossing
is given by $N\simeq\int \frac{H\Upsilon}{|V'|}\,d\phi$. Although this
approximation is valid in the strong dissipative regime
(i.e. $Q\gg1$), by solving numerically the inflationary stage we find
that this formula still constitutes a good approximation to the number
of $e$-foldings for $Q_*\gtrsim 2$. Writing $H\simeq
H_0\,f^{1/2}$ (which holds until the end of inflation) and using Eq.~(\ref{eq28}), the amount of expansion after the observable Universe exits the horizon is given by
\[\label{eq42}
N_*\simeq\frac{(f\textrm{'}_e)^{8/7}\left(1-\delta
_e\right)^{8/7}}{2^{16/7}f_e^{9/7}}\int_{\delta _e}^{\delta
_*}\!\!\frac{f^{2/7}\,d\delta}{(1-\delta
)^{8/7}(f\textrm{'})^{1/7}}\,.
\]

Keeping fixed the number of $e$-foldings $N_*$ it is possible to solve for $\delta_*$ in terms of $\delta_e$ by integrating numerically the above equation. Our strategy now consists of employing this parametrisation $\delta_*=\delta_*(\delta_e)$ and Eq.~(\ref{eq25}) to express the observable quantities in terms of $\delta_e$.

In the absence of the corrected part of the perturbation spectrum, i.e. ${\cal P}_\calr^{(c)}=1$, the spectral index of the perturbation spectrum, defined as
$n_s-1=\frac{d\ln{\cal P_R}}{d\ln k}$, is given by \cite{BasteroGil:2004tg}
\[
\widetilde{n_s}-1\simeq\frac{1}{1+Q_*}
\left(-(2-5A)\epsilon_\phi-3A\eta_\phi+(2+4A)\sigma_\phi\right)\,,
\]
where
\[
\sigma_\phi=m_P^2\frac{V'}{V\phi}\quad\textrm{and}\quad
A=\frac{Q}{1+7Q}\,.
\]
Using the slow-roll equations, the ratio Q at the time of horizon crossing is given by
may be rewritten as follows
\[
Q_*\simeq\frac{C_r^{-3/7}(f\textrm{'}_*)^{6/7}C_\phi^{4/7}}{2^{6/7}3^{1/7}
(1-\delta_*)^{8/7}f_*^{5/7}}\left(\frac{m_P}{\phi_v}\right)^2
\left(\frac{H_0}{m_P}\right)^{2/7}\,.
\]

Upon including the correction ${\cal P}_\calr^{(c)}$,  the spectral index of the corrected spectrum is given by
\[
n_s=\widetilde{n_s}+\frac{d\ln{\cal P}_\calr^{\rm (c)}}{d\ln\,k}\,,
\]
where
\[
\frac{d\ln{\cal P}_\calr^{\rm (c)}}{d\ln\,k}=\frac{d\ln{\cal P}_\calr^{\rm (c)}}{dQ}\,A(10\epsilon_\phi-6\eta_\phi+8\sigma_\phi)\,.
\]

Using the combined WMAP+BAO+SN data and for negligible tensor perturbations, current observations constrain the spectral index to
\mbox{$n=0.963 \pm 0.014$} at the $1$-$\sigma$ level
\cite{Komatsu:2010fb}.

Also interesting for observational purposes is the running of the
spectral index, defined as
\[
n_s'=\widetilde{n_s}'+n_s'^{\rm (c)}\,,
\]
where $\widetilde{n_s}'$ can be written in terms of slow-roll parameters \cite{BasteroGil:2009ec} and
\[\nonumber
n_s'^{\rm (c)}=\frac{d^2\ln{\cal P}_\calr^{\rm (c)}}{d\ln k^{\,2}}=\frac{d^2\ln{\cal P}_\calr^{\rm (c)}}{dQ^2}(10\epsilon_\phi-6\eta_\phi+8\sigma_\phi)^2A^2
\]
\[
+\frac{d\ln{\cal P}_\calr^{\rm (c)}}{dQ}\left[\frac{(10\epsilon_\phi-6\eta_\phi+8\sigma_\phi)^2}
{(1+7Q)^2}+10\epsilon_\phi-6\eta_\phi+8\sigma_\phi\right]A\,.
\]

In the absence of tensor perturbations and using the combined
WMAP+BAO+SN data, the running is constrained to the interval $n'_s =
-0.034 \pm 0.026$ at the $1$-$\sigma$ level \cite{Komatsu:2010fb}.

%%%%%%%%%%%%%%%%%%%%%%%%%%%%%%%%%%%%%%%%%%%%%%%%%%%%%%%%%%%%%%%%%%%%%%
\section{Constraints on gravitino production}
\label{sec.gravitino}
%%%%%%%%%%%%%%%%%%%%%%%%%%%%%%%%%%%%%%%%%%%%%%%%%%%%%%%%%%%%%%%%%%%%%%

Here we discuss  production processes of gravitino and observational
constraints on its abundance. Fist of all, the gravitino is produced
through the thermal scattering among the standard  particles   such as
quarks and gluons. Then the yield value, $Y_{3/2} \equiv n_{3/2} /s$
of the thermally-produced gravitino  is approximately given
by~\cite{Bolz:2000fu,Kawasaki:2004qu,Pradler:2006qh,Rychkov:2007uq,
Kawasaki:2008qe}
\begin{eqnarray}
    \label{eq:thermal}
    Y_{3/2} \simeq  2 \times 10^{-14} \left( \frac{T_{R}}{10^{8} {\rm
    GeV}} \right) \left( 1 + \frac{m_{\tilde{g}}^{2}}{3 m_{3/2}^{2}} \right)
\end{eqnarray}
with $m_{\tilde{g}}$ being the gluino mass. Note that this yield value
is proportional to the reheating temperature.  For  massive unstable
gravitinos with $m_{3/2} \gg 10^{2} $GeV $\sim {\cal
O}(m_{\tilde{g}})$,  the second term in the second bracket  is
negligible.  The energetic products by the decay of gravitino are so
dangerous that the light element abundances are modified by their
annihilations and productions through  scattering processes. When we
consider the case of $m_{3/2} \sim {\cal O}(1)$ TeV, to agree with the
observational light element abundances  we get upper bounds on the
gravitino abundance to be $Y_{3/2} \lesssim 10^{-16}$ and $Y_{3/2}
\lesssim 10^{-14}$ for a branching ratio decaying into hadrons, $B_{h}
= 1$ and $B_{h} = 10^{-3}$,
respectively~\cite{Kawasaki:2004qu,Kawasaki:2004yh}.  The latter value
of $B_{h}$ is a reasonable lower limit on the hadronic branching
ratio~\cite{Kawasaki:2004qu}.  Then we obtain the upper bound on the
reheating temperature in turn through Eq.~(\ref{eq:thermal}) to be
$T_{R} \lesssim {\cal O}(10^{6})$~GeV ($T \lesssim {\cal
O}(10^{8})$~GeV) with a hadronic branching ratio $B_{h} = 1$ ($B_{h} =
10^{-3}$)~\cite{Kawasaki:2004yh,Kawasaki:2004qu} (see Fig.~44 and
Fig.~45 of Ref.~\cite{Kawasaki:2004qu}).

On the other hand, for much larger masses, $m_{3/2}\gtrsim 70 $ TeV with
$B_{h} = 1$ (or $m_{3/2}\gtrsim 20 $TeV with $B_{h} = 10^{-3}$), we
have another type of  upper bound on the abundance of gravitino
which produces the Lightest SUSY Particle (LSP) dark matter as a decay
product. This is estimated to be
% $Y_{3/2} = Y_{\rm LSP} \lesssim 4 \times 10^{-12}
% (\Omega_{\rm LSP}/0.12) (m_{\rm LSP}/10^{2} {\rm GeV})^{-1} $.
%%
\begin{eqnarray}
    \label{eq:YLSP}
    Y_{3/2} = Y_{\rm LSP} \lesssim 4 \times 10^{-12}
    \left( \frac{\Omega_{\rm LSP} h^{2}}{0.12} \right)
    \left( \frac{m_{\rm LSP}}{10^{2} {\rm GeV} } \right)^{-1}.
\end{eqnarray}
By adopting the upper bound on the density parameter of LSP,
$\Omega_{\rm LSP} h^{2} \lesssim 0.12$~\cite{Komatsu:2010fb} with $h$
the reduced Hubble parameter, and a typical scale of the LSP mass
($m_{\rm LSP} \sim 10^{2}$ GeV), this  gives us the upper bound on the
reheating temperature $T_{R} \lesssim 2 \times 10^{10}$ GeV.  We see
that this is much milder than that of BBN  for the
thermally-produced gravitinos with $m_{3/2} \sim 1$ TeV.

Other attractive process to produce gravitino would be the
nonthermal production by the inflation decay through $\phi \to 2
\psi_{\mu}$~\cite{Nakamura:2006uc,Endo:2006zj,Asaka:2006bv,Dine:2006ii,
Endo:2006tf,Kawasaki:2006hm,Endo:2006qk,Kawasaki:2006mb,Endo:2007ih,
Takahashi:2007gw,Endo:2007sz} where  $\psi_{\mu}$ means the
gravitino. Since we expect a relatively low reheating temperature for
the second reheating, this process can importantly affect the result
because the abundance of the nonthermally-produced gravitino is
inversely proportional to the reheating temperature, $Y_{3/2} \sim 2
B_{3/2} Y_{\phi} \sim 3 M_{P} \Gamma_{\phi \to 2 \psi_{\mu}}  /
2m_{\phi} T_{R}  $ where $\Gamma_{\phi \to 2 \psi_{\mu}}$ is the
differential decay width to a pair  of gravitino, $B_{3/2} \equiv
\Gamma_{\phi \to 2 \psi_{\mu}} / \Gamma_{\phi}$ is the branching ratio
into a pair of gravitinos, and $\Gamma_{\phi}$ is the total decay rate
of the inflaton filed ($\simeq T_{R}^{2}/ M_{p}$). Adopting a result
of Ref.~\cite{Endo:2007sz}, the yield of the nonthermally-produced
gravitino is estimated to be
\begin{eqnarray}
    \label{eq:nonthermal}
    Y_{3/2} &\sim& 7 \times 10^{-15} \nonumber \\
    &\times&
    \left(\frac{T_{R}}{10^{8} {\rm GeV}}\right)^{-1}
    \left(\frac{\langle \phi \rangle}{10^{18} {\rm GeV}} \right)^{2}
    \left(\frac{m_{\phi}}{10^{8} {\rm GeV}} \right)^{2}, \nonumber
    \\
\end{eqnarray}
in case of $m_{\phi} < \sqrt{m_{3/2} m_{p}}$ where $\langle \phi
\rangle$ and $m_{\phi}$ are  the vev and the mass of the inflaton at
the decay epoch. Therefore the yield value explicitly depends on the
model parameters such as $\langle \phi \rangle$ and $m_{\phi}$, which
means that we need to analyse the effect for every models. We will
also discuss this type of the constraint on some models simultaneously
in the following sections.

% As will be discussed after
% (\ref{eq:H_0}), the inflaton masses would be $10^{8} {\rm GeV} \lesssim
% m_{\phi}$ in this model because $m_{\phi} \sim m$ at the oscillating
% phase. Then we get $T_{R} \gtrsim 10^{8}$ GeV ($T_{R} \gtrsim 10^{10}$
% GeV) for $B_{h} = 10^{-3}$ ($B_{h} =1 $).

%%%%%%%%%%%%%%%%%%%%%%%%%%%%%%%%%%%%%%%%%%%%%%%%%%%%%%%%%%%%%%%%%%%%%%
\section{Some examples}\label{sec.examp}
%%%%%%%%%%%%%%%%%%%%%%%%%%%%%%%%%%%%%%%%%%%%%%%%%%%%%%%%%%%%%%%%%%%%%%

In this section we illustrate the procedure developed in the
previous section in order to compute the second reheating
temperature $T_{R_2}$ in some simple models.

\subsection{Tree level potential}\label{sec.tl}
We examine a simple extension of the hilltop model studied in
Ref.~\cite{BuenoSanchez:2008nc} where we consider an additional
quartic term
\[\label{eq3}
V=V_0-\frac12|m^2|\phi^2+\frac{\lambda}{4!}\,\phi^4\,,
\]
that stabilises the potential and provides a minimum to support
the field oscillations after inflation. The parameter $\lambda$ is
tuned so that the scalar potential vanishes at the minimum.
With this choice, the
scalar potential is written as in Eq.~(\ref{eq26}) with
\[\label{eq1}
f=\delta^2(2-\delta)^2\,,
\]
and defining $|\eta_0|\equiv|m^2|/3H_0^2$ the expectation value of
the field is $\phi_v=2|\eta_0|^{-1/2}m_P$.

\begin{center}\begin{figure}[tbp]
\epsfig{file=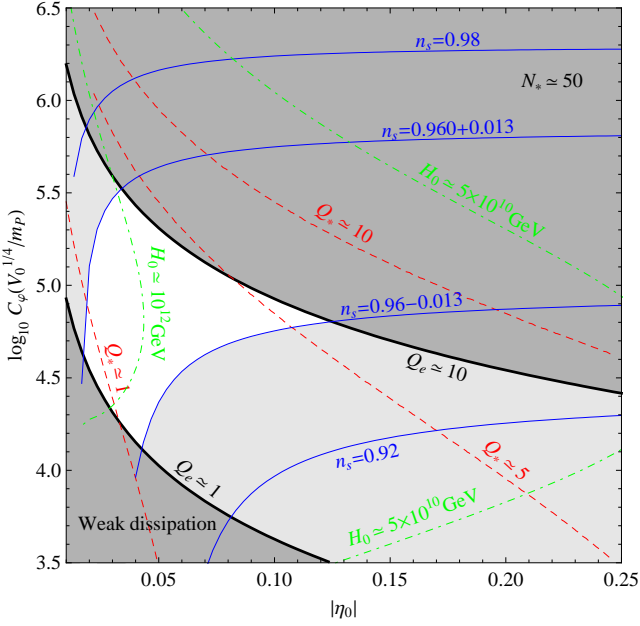,width=8.5cm}\caption{Region of the
parameter space allowed by observations. Heavy-shaded areas are
excluded either because the inflaton field decays completely after
inflation ($Q_e>10$) or because the system is in the weak dissipation
regime ($Q \lesssim 1$). The light-shaded area identifies the region
where the inflaton field does not decay completely after inflation
$(1\lesssim Q_e\lesssim10)$ and the spectral index is outside its
$95\%$ CL interval. In the unshaded region the inflaton avoids its
complete decay after inflation and the spectral index is within $95\%$
CL. Dashed lines (red) correspond to constant values of the ratio
$Q_*$, thin lines (blue) correspond to constant values of $n_s$ and
dot-dashed lines (green) correspond to constant $H_0$. }\label{fig1}
\end{figure}
\end{center}

In order to display the typical range of parameters allowed by
observations (see Fig.~\ref{fig1}) we fix the amount of inflation
after horizon exit to the typical value $N_*\simeq50$. Using the
parametrisation $\delta_*=\delta_*(\delta_e)$ obtained from
Eq.~(\ref{eq42}), the spectral index and its running are determined by
$\delta_e$ and $\eta_0$. We use Eq.~(\ref{eq28}) to plot our results
in Fig.~\ref{fig1} in the plane $\eta_0$-$\log_{10}C_\phi
(V_0^{1/4}/m_P)$. In the 1-$\sigma$ window of
$n_s$ displayed in Fig.~\ref{fig1} we find
\[
10^{10}\,{\rm GeV}\lesssim H_0 \lesssim 10^{12}\,{\rm
GeV}\,,
\label{eq:H_0}
\]
which corresponds to the range $10^7\lesssim
C_\phi\lesssim10^8$ and $10^8 {\rm GeV} \lesssim
m \lesssim10^{10} {\rm GeV}$. Using Eq.~(\ref{eq39}) to compute the first
reheating temperature $T_{R_1}$, we find that this ranges between
\[\label{eq13}
10^{13}\,{\rm GeV}\lesssim T_{R_1}\lesssim10^{14}\,{\rm GeV}\,.
\]
This is then far in excess of the current bounds for thermal gravitino
production~\cite{Kawasaki:2004yh,Kawasaki:2004qu}. Also, in the entire space depicted in
Fig.~\ref{fig1} the running of the spectral index ranges between
$-10^{-3}\lesssim n_s'\lesssim -10^{-4}$.

\begin{figure}[htbp]
\epsfig{file=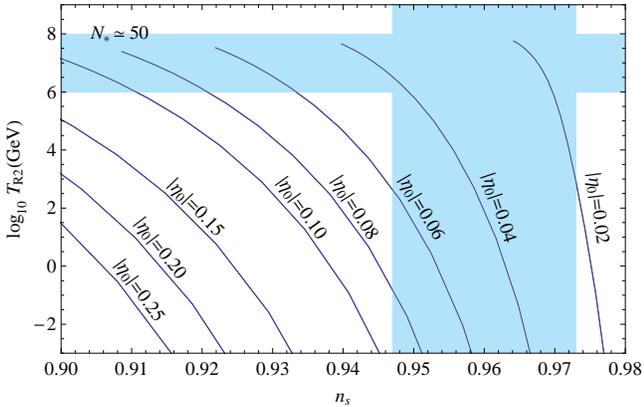,width=8.5cm}
\caption{Reheating temperature $T_{R_2}$ vs
spectral index $n_s$ for several values of $|\eta_0|$. We set
set $m_{3/2}\sim{\rm TeV}$ and $B_h=10^{-3}$ (see Fig.~\ref{fig3}). Only the part
of the curves  compatible with strong dissipation is shown. The
horizontal shaded band corresponds to the range where the problem of
thermal overproduction of gravitinos is less
severe \cite{Kawasaki:2004yh,Kawasaki:2004qu} The vertical band
corresponds to the preferred range of the spectral index at
1-$\sigma$.}\label{fig2}
\end{figure}

Before estimating $T_{R_2}$ by using Eq.~(\ref{eq15}), we need to
obtain an accurate estimate of the average scalar density
$\langle\rho_\phi\rangle$ at late times. As mentioned before, the
estimated $T_{R_2}$ in Eq.~(\ref{eq15}) is not accurate because of
two reasons. The first one owes to have assumed that the field
undergoes a sudden transition to the phase of fast oscillations,
whereas the second one arises because the condition that determines
the onset of the fast-roll motion itself involves an order of
magnitude estimate [cf. Eq.~(\ref{eq19})]. As a result, the redshift
factor $(a_o/a_e)$ in Eq.~(\ref{eq32}) cannot be computed accurately. Using
Eq.~(\ref{eq31}) and writing $H_o\simeq H_e(a_e/a_o)^2$ we obtain
that $Q_o$ is given by
\[
Q_o\simeq
Q_e\frac{(1-\delta_e)^2}{(1-\delta_o)^2}\left(\frac{a_e}{a_o}\right)\,.
\]
Consequently, the ratio $Q_o$ can be determined only up to a
factor of order unity, which can change the reheating temperature
$T_{R_2}$ substantially owing to its exponential dependence on
$Q_o$ [cf. Eq.~(\ref{eq14})]. Note also that
although $\delta_o$, obtained by solving numerically
Eq.~(\ref{eq8}), cannot be accurately computed either, this does
not imply a significant error in $Q_o$ because $\delta_o\ll1$ in
any case.

\begin{center}
\begin{figure}[ht]
\epsfig{file=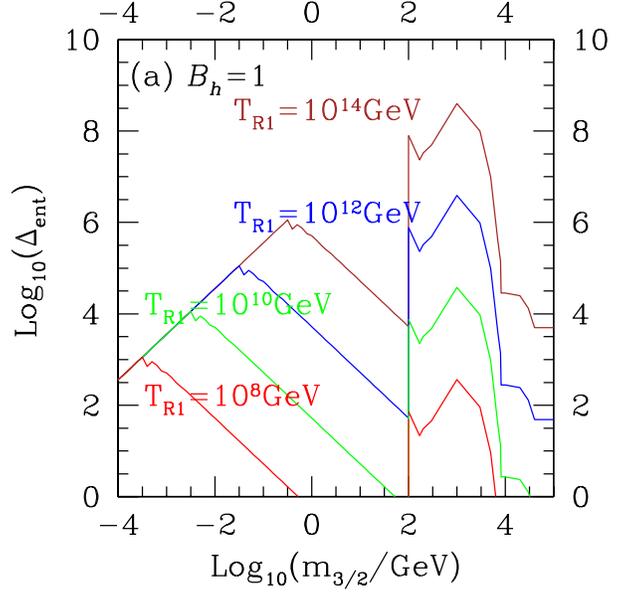,width=9.5cm}\\
\vspace{-1.3cm}
\epsfig{file=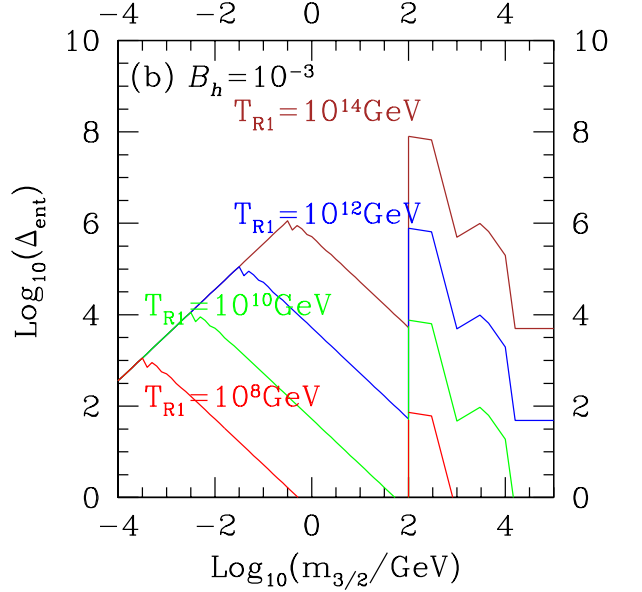,width=9.5cm}
\caption{ Plot of $\Delta_{\rm
ent} = T_{\rm eq}/T_{\rm R_{2}}$ for  a hadronic branching ratio, (a)
$B_{h}=1$ and (b) $B_{h}=10^{-3}$, which is necessary for sufficient
dilution of the gravitino abundance after the second reheating  to
agree with the observational constraints.  From the top to the bottom,
we plot the cases of $T_{R_{1}} = 10^{14},10^{12}, 10^{10}$ and $10^{8}$
GeV, respectively. }
\label{fig3}
\end{figure}
\end{center}

In the Appendix we show that in order to obtain an accurate
estimate for $T_{R_2}$ it suffices to perform the transformation
\[
Q_o\to \widetilde{Q}_o=\alpha Q_o\,,
\]
where $\alpha\simeq1.7$.
Applying such a transformation, we use the late-time corrected
densities in Eqs.~(\ref{eq40}) and (\ref{eq34}) to estimate
$T_{R_2}$ from Eq.~(\ref{eq15})
\[
T_{R_2}\sim2C_r^{-1/4}V_0^{1/4}\frac{e^{-3\widetilde{Q}_o}f_o}
{f_e^{3/4}}\left(\frac{a_o}{a_e}\right)^3\left( \frac{a_{\rm
eq}}{a_{\rm dec}}\right)^{3/4}\,.
\]
The decay of the inflaton field must produce enough entropy as to
dilute the excess of thermal gravitinos produced at temperature
$T_{R_1}$. Given a certain reheating temperature $T_{R_1}$ and for a
certain gravitino mass $m_{3/2}$, this is achieved provided the second
reheating temperature is not larger than the upper bound discussed in
Sec.~\ref{sec.gravitino},
\[\label{eq45}
T_{R_2}=\Delta^{-1}_{\rm ent}(m_{3/2},T_{R_1})T_{\rm eq}\,.
\]
Here $\Delta_{\rm ent}$ is the entropy dilution factor and $T_{\rm
eq}$ is the temperature of the radiation bath at the time of
matter-radiation equality, which is  the  same  order of
$T_{R_{1}}$. In Fig.~\ref{fig2} we present our results for $T_{R_2}$ vs $n_s$
for several values of $|\eta_0|$ and for the typical values
\mbox{$N_*\simeq50$}. To obtain the curves in Fig.~\ref{fig2} we
set $m_{3/2}\sim{\cal O}({\rm TeV})$ and $B_h=10^{-3}$, hence
$\Delta_{\rm ent}\sim10^6(T_{R_1}/10^{14}{\rm GeV})$. Furthermore, we
arrange the complete decay of the inflaton at the earliest time
compatible with the dilution of the gravitino overproduction, i.e.
we set $(a_{\rm eq}/a_{\rm dec})^{3/4}=\Delta_{\rm ent}^{-1}$
[cf. Eqs.~(\ref{eq15}) and (\ref{eq45})]. If
the decay of the inflaton is further delayed then $T_{R_2}$
decreases accordingly. Note also that an excessively
low reheating temperature may pose a problem when the
non-thermal production of gravitinos is considered, therefore
it is desirable that the inflaton field decays ``soon'' enough to
obtain a sufficiently large reheating temperature.

The shaded areas displayed in Fig.~\ref{fig2} enclose the 1-$\sigma$
window of $n_s$ and the range of temperatures where the thermal
gravitino overproduction are typically less problematic, i.e.
\mbox{$T_{R_2}\sim10^6\,$GeV} for $B_{h} =1$ and
\mbox{$T_{R_2}\sim10^8\,$GeV} for $B_{h}
=10^{-3}$~\cite{Kawasaki:2004yh,Kawasaki:2004qu}. Our plots make clear
that, provided $|\eta_0|$ lies in the range
$0.02\lesssim|\eta_0|\lesssim0.05$, the inflaton, while giving rise to
a thermal perturbation spectrum in agreement with current
observations, manages to drive a late matter-dominated epoch and give
rise to a radiation bath with temperature $T_{R_2} \lesssim 10^6 -
10^8\,$GeV after its complete decay. This is an interesting result,
and hence it is worth mentioning that the  ``survival" of the inflaton
field owes to the moderate growth of $Q$ during inflation. For the
hilltop potential in Eq. (\ref{eq3}), in the slow-roll regime, the
dissipative ratio evolves as \cite{BasteroGil:2009ec}: \be \frac{d \ln
Q}{d N_e} \simeq - \frac{|\eta_0|}{1 + 7Q} \left( 2 - 5 |\eta_0|
  \left(\frac{\phi}{m_P}\right)^2 \right) \,, \ee and thus $Q$ starts
decreasing and increases only towards the end of inflation when
$\phi/m_P\sim |\eta_0|^{-1/2}$. In particular, we find $1\lesssim
Q_e/Q_*\lesssim 3.5$ within the allowed region displayed in
Fig.~\ref{fig1}. Ultimately, the fact  that $\Upsilon$ hardly grows
during inflation is due to the moderate steepness of the
potential. Therefore, to avoid the complete decay of the inflaton it
is only necessary to tune the model parameters so that the last stage
of inflation takes place with the system not far away from the weak
dissipation limit, and therefore for not too large values of
$|\eta_0|$.  We note however that for the range
$0.05\lesssim|\eta_0|\lesssim0.10$ where the gravitino overproduction
is avoided, the vacuum expectation value of the field
$\phi_v=2|\eta_0|^{-1/2}m_P$ is slightly above the Planck scale.

In Fig.~\ref{fig3} $\Delta_{\rm ent}$ is plotted vs
$m_{3/2}$ for several values of $T_{R_1}$ for (a) $B_{h}$ = 1 and (b)
$B_{h}=10^{-3}$. For $m_{3/2} \ge 10^{2}$ GeV, we adopted the
constraints from the unstable gravitinos~\cite{Kawasaki:2004qu}. On
the other hand, for $m_{3/2} < 10^{2}$ GeV, we adopted the upper bound
on the density of the stable gravitino not to exceed the LSP density
shown in Eq.~(\ref{eq:YLSP})  by replacing $m_{\rm LSP}$ to $m_{3/2}$,
and using Eq.~(\ref{eq:thermal}). Then the second term in the second bracket
of Eq.~(\ref{eq:thermal}) dominates.  The left-side tail  with respect
to the peak structure in Fig.~\ref{fig3} comes from the saturation of
the number density due to the thermalisation of gravitino, $Y_{3/2}
\lesssim 1/g_{*}^{3/2}$. In  the current cases where
$T_{R_1}\lesssim10^{14}\,{\rm GeV}$ and $m_{3/2}={\cal O}({\rm TeV})$,
the required dilution factors are of the order of $\Delta_{\rm
ent}\gtrsim {\cal O}(10^{8})$  and $\gtrsim {\cal O}(10^{6})$ for
$B_{h}$ = 1 and $10^{-3}$, which implies the constraint
$T_{R_2}\lesssim{\cal O}(10^6)\,$GeV and $T_{R_2}\lesssim{\cal
O}(10^8)\,$GeV, respectively.

From (\ref{eq:H_0}) and $m \sim \sqrt{|\eta_{0}|} H_{0}$,  we find
that the inflaton masses would be  in the range of $10^{8} {\rm GeV}
\lesssim m_{\phi} \lesssim 10^{10} {\rm GeV}$ because $m_{\phi} \sim
m$ at the oscillating epoch. By considering the constraint from the
nonthermal production of the gravitino given in
Eq.~(\ref{eq:nonthermal}) with $\langle \phi \rangle \sim m_{p}$ and
$m_{\phi} \sim 10^{8}$~GeV, we get $T_{R_{2}} \gtrsim 10^{8}$ GeV
($T_{R_{2}} \gtrsim 10^{10}$ GeV) for $B_{h} = 10^{-3}$ ($B_{h} =1
$). It is attractive that the gravitino abundances produced both
thermally and nonthermally  agree with any observational constraints
if we adopt the hadronic branching ratio to be $B_{h} = 10^{-3}$.

The strong dissipation limit of warm inflation is known to result
in the generation of Non-Gaussian effects
\cite{Gupta:2002kn,Moss:2007cv,Chen:2007gd}, and several
models of warm inflation have been constructed with such
effects \cite{BuenoSanchez:2008nc,BasteroGil:2009ec,Matsuda:2009eq}.
In such a limit it
was shown in Ref.~\cite{Moss:2007cv} that entropy fluctuations
during warm inflation play an important role in generating
non-Gaussianity, with the prediction
\[
-15\ln\left(1+ \frac{Q_*}{14}\right)-\frac{5}{2}\lesssim f_{\rm
NL} \lesssim \frac{33}{2} \ln
\left(1+\frac{Q_*}{14}\right)-\frac{5}{2}\,.
\]
To estimate the magnitude of $f_{\rm NL}$ for this model within the
allowed region shown in Fig.~\ref{fig2}, it is enough to compare the
results plotted there (left-hand panel) with Fig.~\ref{fig2}. We then
see that in the allowed region $Q_*$ is at most of order 10, which
corresponds to $|f_{\rm NL}|\lesssim10$. This is well within the
observed range of the $f_{\rm NL} = 32 \pm 21~(68\% {\rm
C.L.})$~\cite{Komatsu:2010fb}. Note that the PLANCK satellite
\cite{planck}, which was launched recently, will be sensitive to
non-Gaussianity at the level $|\Delta f_{\rm NL}|={\cal
O}(5)$~\cite{planck,PLANCK_coll}.

\subsection{Supergravity inspired model}\label{sec.sugra}
%In order to keep the vacuum expectation value of the field $\phi_v$
%below the Planck scale
We consider now a model typical of supergravity theories
\cite{Izawa:1996dv,Izawa:1997df} as an example. In this kind of
models, the inflaton superfield $\Phi$ is assumed to have an $R$
charge $2/(n+1)$ allowing the superpotential
\[
W_0=-\frac{g}{n+1}\,\Phi^{n+1}\,,
\]
with $n$ positive and $g$ a coupling constant.
The continuous $U(1)_R$ symmetry is assumed to be dynamically
broken to a discrete $Z_{2nR}$ at a scale $v\ll m_P$ generating the
superpotential
\[
W_{\rm eff}=v^2\Phi-\frac{g}{n+1}\,\Phi^{n+1}\,.
\]
The inflaton field
$\phi(x)/\sqrt{2}$ is then identified as the real part of the
scalar component of the superfield $\Phi$. Taking
the $R$-invariant  K\"ahler potential,
$K=|\Phi|^2+ k |\Phi|^4/4 + \cdots$, the scalar potential is given by
\[\label{eq41}
V(\phi)\simeq
v^4-\frac{k}{2}v^4 \frac{\phi^2}{m_P^2}-\frac{g}{2^{n/2-1}}v^2 \frac{\phi^n}{m_P^{n-2}}
+\frac{g^2}{2^n}\frac{\phi^{2n}}{m_P^{2n-4}}\,,
\]
which in the absence of dissipation was shown to be flat enough to
support inflation and  generate the appropriate perturbation spectrum.
At the vacuum, one has
\[
\phi_v\simeq\sqrt{2}\left(\frac{v^2}{m_P^2g}\right)^{1/n} m_P\,,
\]
and the scalar potential is negative due to the contribution proportional
to $k$ from the non-minimal K\"ahler potential. In the original model
\cite{Izawa:1996dv} that was canceled by a positive SUSY breaking
effect contributing an amount $\Lambda_{\rm SUSY}^4$, which then fixes
the gravitino mass. Thus, the main role of the non-minimal K\"ahler $k$
term was to fix the scale for the gravitino mass. On the other hand,
by considering $k \sim g$ and $v \ll m_P$, one has in general $v \ll
\phi_v$, and the quadratic term in Eq. (\ref{eq41}) can be neglected
whenever $\phi \sim \phi_v$. This is what we expect during the last
50-60 e-folds of warm
inflation. As discussed earlier, warm inflation finishes giving way to
a radiation dominated Universe when $\phi \lesssim \phi_v$, while the
field is still in slow-roll. Owing to the slow-roll motion, it is
natural to expect that $v \ll \phi \lesssim \phi_v$. Consequently, the
scalar potential during the last stage of inflation can be
approximated by neglecting the quadratic contribution due to the
coupling $k$.

%\[\label{eq44}
%m_{3/2}\simeq\frac{n}{n+1}\left(\frac{v}{m_P}\right)^2
%\left(\frac{v^2}{m_P^2g}\right)^{1/n} m_P\,.
%\]

In the following we will just set $k=0$ to discuss this kind of scalar
potentials in the context of warm inflation. This also means that we do
not consider any particular susy breaking mechanism, neither we link
the gravitino mass to any particular vacuum scale.
The potential Eq.~(\ref{eq41})
can be written as in Eq.~(\ref{eq26}) with $V_0=v^4$ and
\[
f(\delta)\simeq ( 1 -(1-\delta)^n)^2\,,
\]
and we can apply the procedure developed in the last section to find
the range of model parameters consistent with observations.
For a general power $n$, the dissipative ratio evolves
during inflation as:
\be
\frac{d \ln Q}{d N_e} \simeq - \frac{|\eta_0|/(n-1)}{1 + 7Q} \left( 14-6n - 5
  \frac{|\eta_0|}{n-1} \left(\frac{\phi}{m_P}\right)^n \right) \,.
\ee
When $n=2$ we  recover the potential
studied in the previous subsection, with $|\eta_0|= g v^2/m_P^2$, and
a moderate decrease/increase of $Q$ during inflation. On the other
hand, for steeper potentials with $n > 2$, $Q$ always
increases. Even if the observable universe exits the horizon when the
system is still in the weak dissipative regime (i.e. $Q_*\lesssim 1$), inflation
typically finishes into the strong dissipative regime. We are
interested here on finding the model parameters which lead to  a
moderate value $Q_e\sim  O(10)$ when $n >2$.

\begin{center}
\begin{figure}[tb]
\epsfig{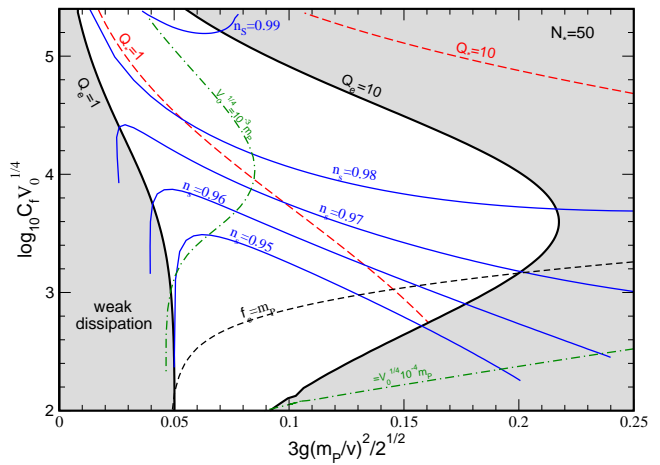}\caption{Parameter space available
  to the strong dissipative regime of warm inflation for the sugra
  potential Eq. (\ref{eq41}), with $n=3$ and $k=0$. In the left shaded
  area the system is in the weak dissipative regime ($Q_e <1$), while
  the right shaded area is excluded because the inflation field decays
  completely after inflation ($Q_e \gtrsim 10$). The dashed lines
  (red) correspond to constant values of $Q_*$, while the thin (blue)
  lines are those of constant $n_s$. The short-dashed line gives that
  of constant field end value $\phi_e$. Below that line
  the field is always subplanckian, and it is $m_P < \phi < 10 m_P$ above. The
  dot-dashed (green) lines are those of constant vacuum energy
  $V_0^{1/4}=v$.
}
\label{fig4}
\end{figure}
\end{center}

Fig.~\ref{fig4} depicts the range of parameters allowed by
observations for $n=3$ and $N_*\simeq50$. In this case the increase of
the dissipative ratio $Q$ is still moderate and we can find values of
parameters which leads to the strong dissipative regime but with $Q_e
\leq 10$, and for which the prediction for the spectral index is within the
observational range. For larger powers $n \geq 4$, owing to the
steepness of the potential, a large ratio $Q$ is needed to drive
slow-roll inflation. As a result, a substantial part of the region
where the spectral index agrees with observations would be excluded by
the bound  $f_{\rm NL}<74$ \cite{Komatsu:2010fb}, implying
$Q_*\lesssim1.2\times10^4$.  And in regards to the
gravitino overproduction, the substantial increase of the ratio $Q$ during the
inflationary stage reduces considerably the parameter space for which
$Q_e \sim O(10)$.  For most of the parameter space, inflation
finishes well into the strong dissipative regime with $Q_e\gg 1$.
But even if we tune the parameters to avoid too large a value of
$Q_e$, the model with $n \geq 4$ predicts a blue spectrum and no
tensors. This is in conflict with observations, which in the absence
of  primordial tensor perturbations favour a red tilted
spectrum. Nevertheless, we remark that the blue spectrum stems from the
large correction ${\cal P}_\calr^{(c)}$ to the perturbation
spectrum. If one disregards such a correction, i.e. ${\cal
  P}_\calr^{(c)}=1$, the model gives rise to a red tilted spectrum
with $0.956\lesssim n_s\lesssim0.974$, in agreement with current
observations. Still, the dissipative mechanism with too step
potentials does not help with the gravitino problem.

\section{Conclusions}\label{sec.concl}

In this paper we have shown that warm inflation models can
lead to a new mechanism for controlling
gravitino overproduction. The residual
oscillating energy of the inflaton field after inflation eventually
dominates the energy density of the universe. Then the late-time
entropy production by its decay can really dilute the gravitinos which
are thermally-produced in the radiation dominated epoch just after
warm inflation ends. Even if we consider the nonthermal production of
gravitino, this scenario is consistent with the observational
constraints. We have demonstrated that this mechanism is applicable to
a large class of models, when the dominant term during inflation goes
like $\phi^n$ with $n=2\,,3$, with a mild decrease/increase of the
dissipative parameter during inflation such that still $Q_e\lesssim
10$ by the end of inflaton. For steeper potentials, the increase of the
dissipative parameter would lead to $Q_e \gg 10$ for most of the
parameter space, and the complete decay of
the inflaton by the end.
%Although we have
%provided a few explicit examples, this paper has not delved into a
%detailed analysis of observational constraint.

We have also discussed the possibility to detect the non-Gaussianity
of the order of $f_{\rm NL} \sim 10$ which originates from the strong
dissipation in the current models of warm inflation. The PLANCK
satellite will be able to detect this signature by which we can
distinguish the current model from  the normal cold inflation models.

\section*{Acknowledgements}
JCBS is supported by the European Research and Training Network
MRTN-CT-2006 035863-1 (UniverseNet), Ministerio de Ciencia e
Innovaci\'on (Spain) through Research Projects No. ESP2007-30785-E, No.
FIS2006-05895 and No. FIS2010-17440, and by Universidad Complutense de
Madrid and Banco Santander through the Grant No. GR58/08-920911.  AB was funded by
STFC. K.K. was partly supported by the Center for the Promotion of
Integrated Sciences (CPIS) of Sokendai, and Grant-in-Aid for
Scientific Research on Priority Areas No. 18071001,  Scientific
Research (A) No.22244030 and  Innovative Areas No.  21111006.
M.B.G. is partially supported by M.E.C. grant FIS2007-63364 and by the
Junta de Andaluc\'{\i}a group FQM101.

\section{Appendix}
In this section we obtain an accurate estimate for $T_{R_2}$. In
order to do so, we need a reference time to compare the predicted
scalar density after inflation, Eq.~(\ref{eq14}), with the value
obtained from the numerical solution of the system,
Eqs.~(\ref{eq21}) and (\ref{eq29}). An appropriate
``checkpoint'' is provided by the time when the system moves into
the weak dissipation regime: $\Upsilon_w=3H_w$ (or $Q_w=1$). This
checkpoint is appropriate because when $\Upsilon=3H$ the field is
already performing fast oscillations, hence the density is well
approximated by the average in Eq.~(\ref{eq14}). Also, for the
cases of interest (when the inflaton does not decay completely
right after inflation), the system is not far away from the weak
dissipation regime. This then allows us to obtain numerically the
scalar density at this time quite easily, as $\Upsilon=3H$ not too
late after the field starts oscillating.

Using Eq.~(\ref{eq14}), the predicted average scalar density when
the system reaches the weak dissipation regime is
\[
\langle\rho_\phi\rangle_w\simeq(\rho_\phi)_o\,Q_o^{-3}\exp\left
\{-3\,Q_o\left(1-Q_o^{-1}\right)\right\}\,,
\]
where $a_o/a_w=Q_o^{-1}$ since $Q$ falls as $a^{-1}$ after
inflation. Because the ratio $Q_o$ is determined up to a factor of
order 1, it is possible to match the above prediction to the
numerical solution at the time $a=a_w$ by the performing the
substitution
\[
Q_o\longrightarrow \widetilde{Q}_o=\alpha\,Q_o\,,
\]
and then finding an appropriate value for $\alpha$. Within the
allowed region displayed in Fig.~\ref{fig1} and fixing
$(\rho_\phi)_o=2V_o$, we find that the average scalar
density is matched to the numerical solution by
taking $\alpha\simeq1.70$. Hence, at late times ($a\gg a_w$) the
average scalar density and the radiation density are well
approximated by [cf. Eq.~(\ref{eq14})]
\[\label{eq40}
\langle\rho_\phi\rangle\simeq(\rho_\phi)_o(\widetilde{Q}_o)^{-3}\left(\frac{a_w}{a}
\right)^3\exp\left\{-3\,\widetilde{Q}_o\right\}\,.
\]
and
\[\label{eq34}
\rho_r\simeq
2(\rho_r)_e(\widetilde{Q}_o)^{-4}\left(\frac{a_e}{a_o}\right)^4\left(\frac{a_w}{a}\right)^4
\,,
\]
with the redshift factor $(a_e/a_o)$ as given by Eq.~(\ref{eq32}).
The factor of 2 in the last equation is necessary so that $\rho_r$
matches its numerical solution at $a=a_w$. Although put by hand,
the introduction of such a factor is justified because after
potential-radiation equality most of the scalar density, i.e. an
amount $V_e=(\rho_r)_e$, is transferred to the radiation bath in a
Hubble time or so.

%%%%%%%%%%%%%%%%%%%%%%%%%%%%%%%%%%%%%%%%%%%%%%%%%%%%%%%%%%%%%%%%%%%%%%
\begin{thebiblio}{99}
%%%%%%%%%%%%%%%%%%%%%%%%%%%%%%%%%%%%%%%%%%%%%%%%%%%%%%%%%%%%%%%%%%%%%%

\bibitem{Fukugita:1986hr}
  M.~Fukugita and T.~Yanagida,
  %``Baryogenesis Without Grand Unification,''
  Phys.\ Lett.\  B {\bf 174}, 45 (1986).
  %%CITATION = PHLTA,B174,45;%%

\bibitem{lepto}
  G.~F.~Giudice, A.~Notari, M.~Raidal, A.~Riotto and A.~Strumia,
  %``Towards a complete theory of thermal leptogenesis in the SM and MSSM,''
  Nucl.\ Phys.\  B {\bf 685} (2004) 89
  [arXiv:hep-ph/0310123];
  W.~Buchmuller, P.~Di Bari and M.~Plumacher,
  %``Leptogenesis for pedestrians,''
  Annals Phys.\  {\bf 315} (2005) 305
  [arXiv:hep-ph/0401240].

\bibitem{ci} A. H. Guth, Phys. Rev. D {\bf 23}, 347 (1981); K.~Sato,
  Mon.\ Not.\ Roy.\ Astron.\ Soc.\  {\bf 195}, 467 (1981);
  %%CITATION = MNRAA,195,467;%%
A. D.  Linde, Phys.\ Lett.\ B {\bf 108}, 389 (1982); A. Albrecht
and P. J. Steinhardt, Phys.\ Rev.\ Lett.\ {\bf 48}, 1220 (1982);
A. Linde, Phys. Lett. B{\bf 129}, 177 (1983).

%\cite{Berera:1995ie}
\bibitem{Berera:1995ie} A.~Berera,
%``Warm Inflation,''
  Phys.\ Rev.\ Lett.\ {\bf 75}, 3218 (1995).
%[arXiv:astro-ph/9509049].
%%CITATION = PRLTA,75,3218;%%

\bibitem{Berera:2008jn}
A.~Berera,
%``The warm inflationary universe,''
Contemp.\ Phys.\  {\bf 47}, 33 (2006).
%[arXiv:0809.4198 [hep-ph]].

%\cite{Berera:2008ar}
\bibitem{Berera:2008ar}
A.~Berera, I.~G.~Moss and R.~O.~Ramos,
%``Warm Inflation and its Microphysical Basis,''
Rept.\ Prog.\ Phys.\  {\bf 72}, 026901 (2009).
%[arXiv:0808.1855 [hep-ph]].
%%CITATION = RPPHA,72,026901;%%

%\cite{BasteroGil:2009ec}
\bibitem{BasteroGil:2009ec}
M.~Bastero-Gil and A.~Berera,
%``Warm inflation model building,''
Int.\ J.\ Mod.\ Phys.\  A {\bf 24}, 2207 (2009).
%[arXiv:0902.0521 [hep-ph]].
%%CITATION = IMPAE,A24,2207;%%

\bibitem{im} I. G. Moss, Phys. Lett. B {\bf 154}, 120 (1985).

%\cite{Berera:1995wh}
\bibitem{Berera:1995wh} A.~Berera and L.~Z.~Fang,
%``Thermally induced density perturbations in the inflation era,''
  Phys.\ Rev.\ Lett.\ {\bf 74}, 1912 (1995).
%[arXiv:astro-ph/9501024].
%%CITATION = ASTRO-PH 9501024;%%

\bibitem{Berera:1999ws}
A.~Berera,
%``Warm inflation at arbitrary adiabaticity: A model, an existence proof  for inflationary dynamics in quantum field theory,''
Nucl.\ Phys.\ B {\bf 585}, 666 (2000).
%[arXiv:hep-ph/9904409].
%%CITATION = HEP-PH 9904409;%%

%\cite{Berera:1998gx}
\bibitem{Berera:1998gx}
A.~Berera, M.~Gleiser and R.~O.~Ramos,
%``Strong dissipative behavior in quantum field theory,''
Phys.\ Rev.\  D {\bf 58}, 123508 (1998).
%[arXiv:hep-ph/9803394].
%%CITATION = PHRVA,D58,123508;%%

\bibitem{Yokoyama:1998ju}
  J.~Yokoyama and A.~D.~Linde,
  %``Is warm inflation possible?,''
  Phys.\ Rev.\  D {\bf 60}, 083509 (1999)
  [arXiv:hep-ph/9809409].
  %%CITATION = PHRVA,D60,083509;%%

\bibitem{Moss:2006gt}
I.~G.~Moss and C.~Xiong,
%``Dissipation coefficients for supersymmetric inflationary models,''
%  arXiv:
hep-ph/0603266.

\bibitem{BasteroGil:2010pb}
M.~Bastero-Gil, A.~Berera and R.~O.~Ramos,
%``Dissipation and Viscosity Coefficients in Quantum Field Theory,''
arXiv:1008.1929 [hep-ph], (2010).
%%CITATION = ARXIV:1008.1929;%%

\bibitem{br} A. Berera and R. O. Ramos,
Phys. Rev. D{\bf 63}, 103509 (2001).

\bibitem{stringwi}
  M.~Bastero-Gil, A.~Berera, J.~B.~Dent and T. W. Kephart,
  %``Towards Realizing Warm Inflation in String Theory,''
  [arXiv:0904.2195 [astro-ph.CO]].

\bibitem{BasteroGil:2006vr}
  M.~Bastero-Gil and A.~Berera,
  %``Warm inflation dynamics in the low temperature regime,''
  Phys.\ Rev.\  D {\bf 76} (2007) 043515.
%  [arXiv:hep-ph/0610343].

\bibitem{Zhang:2009ge}
Y.~Zhang,
%``Warm Inflation with a General Form of the Dissipative Coefficient,''
JCAP {\bf 0903} (2009) 023.
%[arXiv:0903.0685 [hep-ph]].

%\cite{Taylor:2000jw}
\bibitem{Taylor:2000jw}
A.~N.~Taylor and A.~R.~Liddle,
%``Gravitino production in the warm inflationary scenario,''
Phys.\ Rev.\  D {\bf 64}, 023513 (2001).
%[arXiv:astro-ph/0011365].
%%CITATION = PHRVA,D64,023513;%%

\bibitem{BuenoSanchez:2008nc}
  J.~C.~Bueno Sanchez, M.~Bastero-Gil, A.~Berera and K.~Dimopoulos,
  %``Warm hilltop inflation,''
  Phys.\ Rev.\  D {\bf 77} (2008) 123527.
%  [arXiv:0802.4354 [hep-ph]].

\bibitem{RDcond}

M.~Kawasaki, K.~Kohri and N.~Sugiyama,
  %``Cosmological Constraints on Late-time Entropy Production,''
  Phys.\ Rev.\ Lett.\  {\bf 82}, 4168 (1999);
%  [arXiv:astro-ph/9811437].
  %%CITATION = PRLTA,82,4168;%%
  %``MeV-scale reheating temperature and thermalization of neutrino
  %background,''
  Phys.\ Rev.\  D {\bf 62}, 023506 (2000);
%  [arXiv:astro-ph/0002127].
  %%CITATION = PHRVA,D62,023506;%%
S.~Hannestad,
  %``What is the lowest possible reheating temperature?,''
  Phys.\ Rev.\  D {\bf 70}, 043506 (2004);
%  [arXiv:astro-ph/0403291].
  %%CITATION = PHRVA,D70,043506;%%
K.~Ichikawa, M.~Kawasaki and F.~Takahashi,
  %``The oscillation effects on thermalization of the neutrinos in the universe
  %with low reheating temperature,''
  Phys.\ Rev.\  D {\bf 72}, 043522 (2005).
%  [arXiv:astro-ph/0505395].
  %%CITATION = PHRVA,D72,043522;%%

\bibitem{Kamada:2009hy}
  K.~Kamada and J.~Yokoyama,
  %``On the realization of the MSSM inflation,''
  Prog.\ Theor.\ Phys.\  {\bf 122}, 969 (2010)
  [arXiv:0906.3402 [hep-ph]].
  %%CITATION = PTPKA,122,969;%%

\bibitem{BasteroGil:2004tg}
  M.~Bastero-Gil and A.~Berera,
  %``Determining the regimes of cold and warm inflation in the SUSY hybrid
  %model,''
  Phys.\ Rev.\  D {\bf 71} (2005) 063515;
I. G. Moss and X. Xiong, JCAP {\bf 0811} (2009) 023.
%  [arXiv:hep-ph/0411144].

\bibitem{Turner:1983he}
  M.~S.~Turner,
  %``Coherent Scalar Field Oscillations In An Expanding Universe,''
  Phys.\ Rev.\  D {\bf 28} (1983) 1243.

%\cite{Graham:2009bf}
\bibitem{Graham:2009bf}
C.~Graham and I.~G.~Moss,
%``Density fluctuations from warm inflation,''
JCAP {\bf 0907}, 013 (2009).
%[arXiv:0905.3500 [astro-ph.CO]].
%%CITATION = JCAPA,0907,013;%%

\bibitem{mg}
C. Graham and I. G. Moss, private communication.

\bibitem{Komatsu:2010fb}
  E.~Komatsu {\it et al.},
  %``Seven-Year Wilkinson Microwave Anisotropy Probe (WMAP) Observations:
  %Cosmological Interpretation,''
  arXiv:1001.4538 [astro-ph.CO].
  %%CITATION = ARXIV:1001.4538;%%

\bibitem{Bolz:2000fu}
  M.~Bolz, A.~Brandenburg and W.~Buchmuller,
  %``Thermal Production of Gravitinos,''
  Nucl.\ Phys.\  B {\bf 606}, 518 (2001)
  [Erratum-ibid.\  B {\bf 790}, 336 (2008)]
  [arXiv:hep-ph/0012052].
  %%CITATION = NUPHA,B606,518;%%

\bibitem{Kawasaki:2004qu}
  M.~Kawasaki, K.~Kohri and T.~Moroi,
  %``Big-bang nucleosynthesis and hadronic decay of long-lived massive
  %particles,''
  Phys.\ Rev.\  D {\bf 71}, 083502 (2005).
%  [arXiv:astro-ph/0408426].
  %%CITATION = PHRVA,D71,083502;%%

\bibitem{Pradler:2006qh}
J.~Pradler and F.~D.~Steffen,
%``Thermal Gravitino Production and Collider Tests of Leptogenesis,''
Phys.\ Rev.\  D {\bf 75}, 023509 (2007);
%[arXiv:hep-ph/0608344].
%%CITATION = PHRVA,D75,023509;%%
%\bibitem{Pradler:2006hh}
%  J.~Pradler and F.~D.~Steffen,
%``Constraints on the reheating temperature in gravitino dark matter
%scenarios,''
Phys.\ Lett.\  B {\bf 648}, 224 (2007).
%[arXiv:hep-ph/0612291].
%%CITATION = PHLTA,B648,224;%%

\bibitem{Rychkov:2007uq}
  V.~S.~Rychkov and A.~Strumia,
  %``Thermal production of gravitinos,''
  Phys.\ Rev.\  D {\bf 75}, 075011 (2007)
  [arXiv:hep-ph/0701104].
  %%CITATION = PHRVA,D75,075011;%%

\bibitem{Kawasaki:2008qe}
  M.~Kawasaki, K.~Kohri, T.~Moroi and A.~Yotsuyanagi,
  %``Big-Bang Nucleosynthesis and Gravitino,''
  Phys.\ Rev.\  D {\bf 78}, 065011 (2008)
  [arXiv:0804.3745 [hep-ph]].
  %%CITATION = PHRVA,D78,065011;%%

\bibitem{Kawasaki:2004yh}
  M.~Kawasaki, K.~Kohri and T.~Moroi,
  %``Hadronic decay of late-decaying particles and big-bang nucleosynthesis,''
  Phys.\ Lett.\  B {\bf 625}, 7 (2005)
  [arXiv:astro-ph/0402490].
  %%CITATION = PHLTA,B625,7;%%

01;%%

\bibitem{Nakamura:2006uc}
  S.~Nakamura and M.~Yamaguchi,
  %``Gravitino production from heavy moduli decay and cosmological moduli
  %problem revived,''
  Phys.\ Lett.\  B {\bf 638}, 389 (2006)
  [arXiv:hep-ph/0602081].
  %%CITATION = PHLTA,B638,389;%%

\bibitem{Endo:2006zj}
  M.~Endo, K.~Hamaguchi and F.~Takahashi,
  %``Moduli-induced gravitino problem,''
  Phys.\ Rev.\ Lett.\  {\bf 96}, 211301 (2006)
  [arXiv:hep-ph/0602061].
  %%CITATION = PRLTA,96,211301;%%

\bibitem{Asaka:2006bv}
 T.~Asaka, S.~Nakamura and M.~Yamaguchi,
 %``Gravitinos from heavy scalar decay,''
 Phys.\ Rev.\  D {\bf 74}, 023520 (2006)
 [arXiv:hep-ph/0604132].
 %%CITATION = PHRVA,D74,023520;%%

\bibitem{Dine:2006ii}
  M.~Dine, R.~Kitano, A.~Morisse and Y.~Shirman,
  %``Moduli decays and gravitinos,''
  Phys.\ Rev.\  D {\bf 73}, 123518 (2006)
  [arXiv:hep-ph/0604140].
  %%CITATION = PHRVA,D73,123518;%%

\bibitem{Endo:2006tf}
  M.~Endo, K.~Hamaguchi and F.~Takahashi,
  %``Moduli / inflaton mixing with supersymmetry breaking field,''
  Phys.\ Rev.\  D {\bf 74}, 023531 (2006)
  [arXiv:hep-ph/0605091].
  %%CITATION = PHRVA,D74,023531;%%

\bibitem{Kawasaki:2006hm}
  M.~Kawasaki, F.~Takahashi and T.~T.~Yanagida,
  %``The gravitino overproduction problem in inflationary universe,''
  Phys.\ Rev.\  D {\bf 74}, 043519 (2006)
  [arXiv:hep-ph/0605297].
  %%CITATION = PHRVA,D74,043519;%%

\bibitem{Endo:2006qk}
  M.~Endo, M.~Kawasaki, F.~Takahashi and T.~T.~Yanagida,
  %``Inflaton decay through supergravity effects,''
  Phys.\ Lett.\  B {\bf 642}, 518 (2006)
  [arXiv:hep-ph/0607170].
  %%CITATION = PHLTA,B642,518;%%

\bibitem{Kawasaki:2006mb}
  M.~Kawasaki, F.~Takahashi and T.~T.~Yanagida,
  %``The Gravitino-Overproduction Problem in Inflaton Decay,''
  AIP Conf.\ Proc.\  {\bf 903}, 677 (2007)
  [arXiv:hep-ph/0611166].
  %%CITATION = APCPC,903,677;%%

\bibitem{Endo:2007ih}
  M.~Endo, F.~Takahashi and T.~T.~Yanagida,
  %``Anomaly-Induced Inflaton Decay and Gravitino-Overproduction Problem,''
  Phys.\ Lett.\  B {\bf 658}, 236 (2008)
  [arXiv:hep-ph/0701042].
  %%CITATION = PHLTA,B658,236;%%

\bibitem{Takahashi:2007gw}
  F.~Takahashi,
  %``Inflaton Decay in Supergravity and Gravitino Problem,''
  AIP Conf.\ Proc.\  {\bf 957}, 441 (2007)
  [AIP Conf.\ Proc.\  {\bf 1040}, 57 (2008)]
  [arXiv:0709.1786 [hep-ph]].
  %%CITATION = APCPC,1040,57;%%

\bibitem{Endo:2007sz}
  M.~Endo, F.~Takahashi and T.~T.~Yanagida,
  %``Inflaton Decay in Supergravity,''
  Phys.\ Rev.\  D {\bf 76}, 083509 (2007)
  [arXiv:0706.0986 [hep-ph]].
  %%CITATION = PHRVA,D76,083509;%%

\bibitem{Gupta:2002kn}
  S.~Gupta, A.~Berera, A.~F.~Heavens and S.~Matarrese,
  %``Non-Gaussian signatures in the cosmic background radiation from warm
  %inflation,''
  Phys.\ Rev.\  D {\bf 66} (2002) 043510.
%  [arXiv:astro-ph/0205152].
  %%CITATION = PHRVA,D66,043510;%%

\bibitem{Moss:2007cv}
  I.~G.~Moss and C.~Xiong,
  %``Non-gaussianity in fluctuations from warm inflation,''
  JCAP {\bf 0704} (2007) 007.
%  [arXiv:astro-ph/0701302].
  %%CITATION = JCAPA,0704,007;%%

\bibitem{Chen:2007gd}
  B.~Chen, Y.~Wang and W.~Xue,
  %``Inflationary NonGaussianity from Thermal Fluctuations,''
  arXiv:0712.2345 [hep-th].
  %%CITATION = ARXIV:0712.2345;%%

\bibitem{Matsuda:2009eq}
  T.~Matsuda,
  %``Evolution of the curvature perturbations during warm inflation,''
  JCAP {\bf 0906} (2009) 002
  [arXiv:0905.0308 [astro-ph.CO]].

\bibitem{planck}  Planck Surveyor Mission:\\
  http://www.rssd.esa.int/Planck.

\bibitem{PLANCK_coll}
[Planck Collaboration],
%``Planck: The scientific programme,''
arXiv:astro-ph/0604069;
%%CITATION = ASTRO-PH/0604069;%%
%\bibitem{Komatsu:2009kd}
E.~Komatsu {\it et al.},
%``Non-Gaussianity as a Probe of the Physics of the Primordial Universe and
%the Astrophysics of the Low Redshift Universe,''
arXiv:0902.4759 [astro-ph.CO].
%%CITATION = ARXIV:0902.4759;%%

\bibitem{Izawa:1996dv}
  K.~I.~Izawa and T.~Yanagida,
  %``Natural new inflation in broken supergravity,''
  Phys.\ Lett.\  B {\bf 393} (1997) 331;
%  [arXiv:hep-ph/9608359].
  %%CITATION = PHLTA,B393,331;%%

\bibitem{Izawa:1997df}
  K.~I.~Izawa, M.~Kawasaki and T.~Yanagida,
  %``Dynamical tuning of the initial condition for new inflation in
  %supergravity,''
  Phys.\ Lett.\  B {\bf 411} (1997) 249.
%  [arXiv:hep-ph/9707201].
  %%CITATION = PHLTA,B411,249;%%\bibitem{BasteroGil:2009gh}

\end{thebiblio}
\end{document}